\documentclass{article}
\usepackage[utf8]{inputenc}
\usepackage{amsmath}
\usepackage{amssymb}
\usepackage{braket}
\usepackage{graphicx}
\usepackage{color}
\usepackage{ulem}

\newcommand{\beq}[1]{\begin{equation}}
\newcommand{\eeq}{\end{equation}}
\newcommand{\bea}[1]{\begin{eqnarray}}
\newcommand{\eea}{\end{eqnarray}}

\title{Gravitational wave echos from physical black holes}
\author{Yu-Song Cao\footnote{caoyusong15@mails.ucas.ac.cn}$~^{1}$, YanXia Liu\footnote{yxliu-china@ynu.edu.cn}$~^{2}$, Ding-Fang Zeng\footnote{dfzeng@bjut.edu.cn}$~^{1}$}
\date{}

\begin{document}

\maketitle

\noindent $~^{1}$\small{School of Physics and Optoelectronic Engineering, Beijing University of Technology, Beijing, China}

\noindent $~^{2}$\small{School of Physics and Astronomy, Yunnan University, Kunming 650091, PR China}

\begin{abstract}
    Gravitational wave echos from the coalescence of black hole binaries are frequently viewed as signals beyond general relativity or standard model. In this work, we show that these echos are inevitable even in the coalescence of standard general relativity. This is because it is the physical black holes formed through gravitational collapse serve as faithful
description of astro black holes. To all outside probes, only their asymptotic structure before the horizon formation is accessible. Here, we investigate the scattering of a gravitational wave burst on such physical black holes and pay special attention to the echos from them. Our results uncover distinct features of this echo both in time and frequency domains.
\end{abstract}

\section{Introduction}

It is commonly accepted that black holes with central singularity or circular-line-like singularity are the fate of massive bodies evolution under their self gravitation. So static or stable solutions to the Einstein equation such as the Schwarszchild and Kerr-type are taken as representation of black holes in astronomy for granted. However, serious reflection suggests that the situation is more subtle than commonly believed \cite{epjc,2505,D36-2336,D36-2327,D80,L130,D111-3,jcap10-2,cqg38}. This subtlety arises from the time definition dependence of physical description in General relativity. More precisely for black holes, it is a matter of when does the gravitational collapse completes.

\begin{figure}
\begin{center}
\includegraphics[totalheight=59mm]{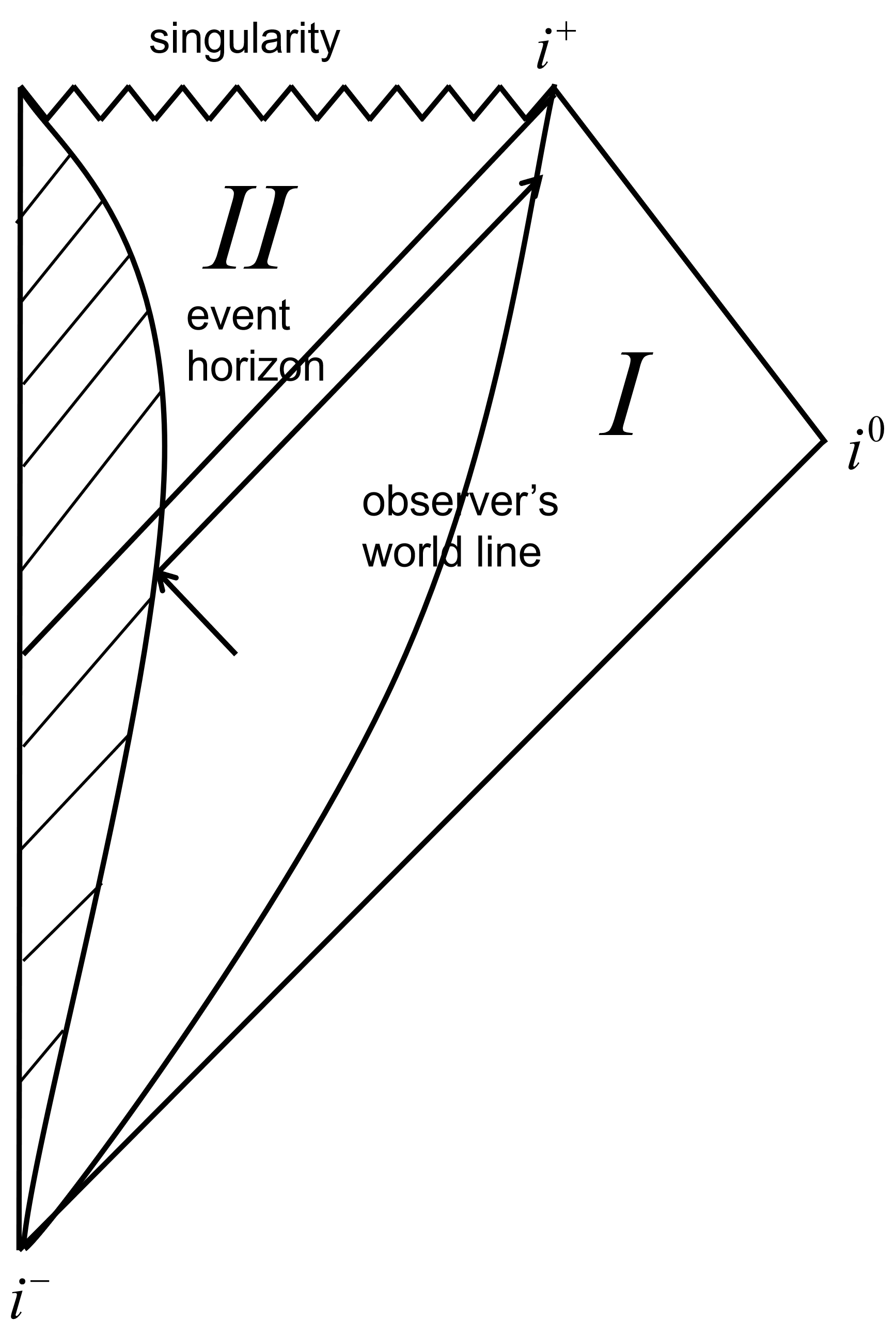}
\caption{\small Penrose diagram of a collapsar, signal and observer. The shaded area is the matter occupied region and the arrows are the world line of the signal. The generation, reflection and detection of the signal all happens in region \textit{I}, where the Schwartzchild time definition is applicable. While Region \textit{II} is outside the Schwartzchild time definition.}
\label{fig:spacetime}
\end{center}
\end{figure}

One can see for example in the Oppenheimer-Snyder model \cite{ro,jose}, all probes living outside the collapsing material will see the surface of the collapsar undergoes an endless contraction towards the event horizon but never reach it at all. To these probes, detectable physical events happen in the Schwarzschild time definition of the collapsar, during which the collapse never finishes, so the horizon never forms in any finite future. While to probes co-moving with the collapsing material, although they will see the surface of the collapsar fall across the event horizon and hit on the singularity in finite duration, what they detect after the horizon formation happens in the domain of Lemaitre time definition but beyond that of the Schwarzschild time definition, is inaccessible to outside probes. To all outside probes, the physics concerning them happen in the Schwarzschild time definition, and the spacetime has an endless imploding and horizonless structure. For illustration one is referred to Fig.(\ref{fig:spacetime}). To emphasize this fact we refer to black holes formed through gravitational collapse as ``physical black holes", while those described by the Schwarzschild or Kerr-metric as ``mathematical black holes". Although many works concerning the black hole phenomenology use mathematic black holes in scenarios where physical black hole is expected, some authors distinguish this two objects carefully and emphasize the necessity in both observational \cite{epjc,2505} and theoretical aspects \cite{D36-2336,D36-2327,D80}.

Since the first gravitational wave signal detected by LIGO ten years ago \cite{ligo}, the world witnessed a surge of growth in gravitational wave astronomy. Now, about a hundred of gravitational wave events have been reported, most of which are from black hole binary coalescence \cite{L1,L2,L3,L4,L5,L6,L7,L8,L9,L10,L11}. With the rapid growth of the gravitational wave antenna sensitivity and the data processing efficiency, in the future, more detailed  structure of the waveform may be detected, among which the gravitational wave echo is a possibility.

Gravitational wave echos from the binary coalescence waveform are widely viewed as a smoking gun for new physics such as quantum black holes \cite{jcap,cqg36,cqg39,D96-2,D101,D111,D111-2,jcap04,L123,L126,D106,D105,D104,D108-2}, exotic compact objects (ECOs) \cite{D95,D107,lrr22,D94,na,D97,D99-2,D99-3,1902,2306,D108}, black hole remnants \cite{D100}, modified gravity theory \cite{D103,D92,D100-3,D100-2,2508,D97-3}, dark matter cloud \cite{epjc84,cpc}, wormhole \cite{D97-2,D109,D101-3,jcap10} and etc. Experimentally, possible echo signals have been reported in black hole coalescence events GW150914, GW151226, and LVT151012 \cite{D96,D99}, while their authenticity are under debate \cite{1612,1701,D98,D97-3,1803}. Currently, more powerful methods are under development aiming at digging echos from the recorded black hole coalescence events \cite{2301}. Theoretically, the echos can be attributed to the boundary condition for gravitational perturbations near the event horizon. In some phenomenological works on ECOs, the Dirichlet boundary condition is applied at the ECO surface for simplicity \cite{jcap}. But to get precise waveforms, the penetration of gravitational wave into the matter region need be considered, on this issue the readers are referred to \cite{D108,D100-4,D101-2,D98-2,D97-4,D98-3,ptep}.

In this work, we focus on the gravitational wave echos from a gravitational burst scattering off a neutral and non-spinning physical black hole. Of course, our method used here has no conceptual difficulty generalizing to the spinning cases. Neglecting the angular momentum is a purely technique simplicity motivated choice. For this black hole, the endless contracting surface serves as a moving reflector modeled by a Dirichlet boundary condition. We observe that the echos from this black hole are fainter than those from ECOs. Their most distinctive feature lies in their frequency content, that is, a huge Doppler red shift between the neighboring echos. Moreover, the time intervals between neighboring echos from are un-equal, they grow as the time increases.

Our paper is organized as follows. In Sec. II we review the definition of physical black holes and the dynamics of its gravitational perturbation. In Sec. III we go through the details of wave packet scattering of physical black holes, where the waveform is presented along with analytical and numerical analysis. Conclusions and discussions are provided in Sec. IV.

Throughout this work the units will be chosen in such a way that $G_{N}=c=1$.

\section{Physical black holes}

Recall in the last section we defined physical black holes as those formed through gravitational collapse. It is very important to remember that in the Schwartzchild time definition where all the probes live and report their measurement, the collapsar undergoes endless contraction towards the event horizon. Here we consider a neutral and non-spinning collapsar with mass $M$, corresponding to the spacetime metric
\begin{equation}\label{eq:metric}
ds^{2}=\left\{
    \begin{aligned}
    &-(1-\frac{2M}{r})dt^{2}+\frac{dr^{2}}{1-\frac{2M}{r}}+r^{2}d\Omega^{2} \quad r>r_{\varepsilon}(t)\\
    &g_{\mu\nu}^{in}dx^{\mu}dx^{\nu} \quad r>r_{\varepsilon}(t),
    \end{aligned}
    \right.
\end{equation}
where $r_{\varepsilon}(t)=2M+\varepsilon(t)$ is the trajectory of the star surface in $\{t,r\}$ plane. Here $\varepsilon(t)$ is small, time varying parameter at the very late stage of the gravitation collapse. Causality dictates that the star surface follows a time-like world line which asymptotically approaches the event horizon so that
\begin{equation}
\frac{dt}{dr_{\varepsilon}(t)}<-\frac{1}{1-\frac{2M}{r_{\varepsilon}(t)}}.
\end{equation}
From Eq.\eqref{eq:metric} we can see the metric outside the matter occupied region is simply vacuum Schwartzchild solution, while the metric $g_{\mu\nu}^{in}$ inside the collapsar is the solution of sourceful Einstein equation
\begin{equation}
G_{\mu\nu}[g^{in}]=8\pi T_{\mu\nu}.
\end{equation}
Here the energy momentum tensor $T_{\mu\nu}$ is that of the matter content of the collapsar. Since the self-gravitation dominates all other interactions and the degeneration pressure during this stage of the collapse, researchers frequently choose neglecting the pressure of collapsing materials and approximating them as pressureless dust, corresponding to the pressureless energy momentum tensor $T_{\mu\nu}=\rho u_{\mu}u_{\nu}$ \cite{epjc,2505,jose,npb917,npb977}. In such cases, the motion of the collapsar surface is freely falling. In this work, we choose not to neglect the pressure so the collapsar surface will undergo an arbitrary ingoing time-like trajectory, where the speed of the surface may be much slower than free falling due to the pressure gradient.

At this point one may wonder since the physical black hole and mathematical black holes are so different, why haven't researchers rule one of them out using accumulated observations? The answer is somewhat surprising. Although astronomers have registered hundreds of black holes from the binary merger process, the existence of event horizon is still yet to be confirmed \cite{1707}. On this issue we recommend the readers to refs.\cite{1707,L116}, where it is pointed out that the ring-down signals are mainly determined by physics of the photon sphere, instead of the physics of event horizon. With identical mass, spin and charge, physical black holes, mathematical black holes and ECOs all have similar ring-down waveforms because all of them have the same photon sphere structure.  In the following part, we will show that the difference between physical and mathematical black holes may be revealed in the post-merger waveforms.

At the ring-down phase of a binary merger process, the spacetime geometry can be decomposed into a background physical black hole plus a small perturbation outside the matter content. The later can be expanded into superpositions of many spherical harmonics whose radial part is governed by the master equation \cite{jcap}
    \begin{equation}\label{eq:eom}
    [\frac{d^{2}}{dt^{2}}-\frac{d^{2}}{dx^{2}}+V_{l}(r)]R_{l}(t,x)=0,
    \end{equation}
    where $x=r+2M\ln (\frac{r}{2M}-1)$ is the tortoise coordinate and
    \begin{equation}\label{eq:potential}
    V_{l}(r)=(1-\frac{2M}{r})[\frac{l(l+1)}{r^{2}}+\frac{(1-s^{2})2M}{r^{3}}]
    \end{equation}
is the centrifuge potential peaked at the photon sphere radius $r=3M$. Here we take $s=2$ for the gravitational wave mode perturbation. For scalar and vector mode perturbations one can simply set $s=0,1$. In the following we will concentrate on the quadrupolar perturbation $l=2$ and write $R_{2}(t,x)$ as $\psi(t,x)$ in the sequel.

To solve the field equation (\ref{eq:eom}) we need to specify the boundary conditions at infinity and on the collapsar surface $r_{\varepsilon}(t)$. The former will be taken as the purely radiational or Sommerfeld condition $(\partial_{t}+\partial_{x})\psi=0$, while the later is assumed to be Dirichlet boundary condition
    \begin{equation}\label{eq:mvb}
    \psi(t,X_{\varepsilon}(t))=0,
    \end{equation}
where $X_{\varepsilon}(t)$ is the trajectory of the collapsar surface expressed in the tortoise coordinate.

\section{Echos from the scattering of wave packets}

\subsection{Analytical analysis}

\begin{figure}
\begin{center}
\includegraphics[totalheight=59mm]{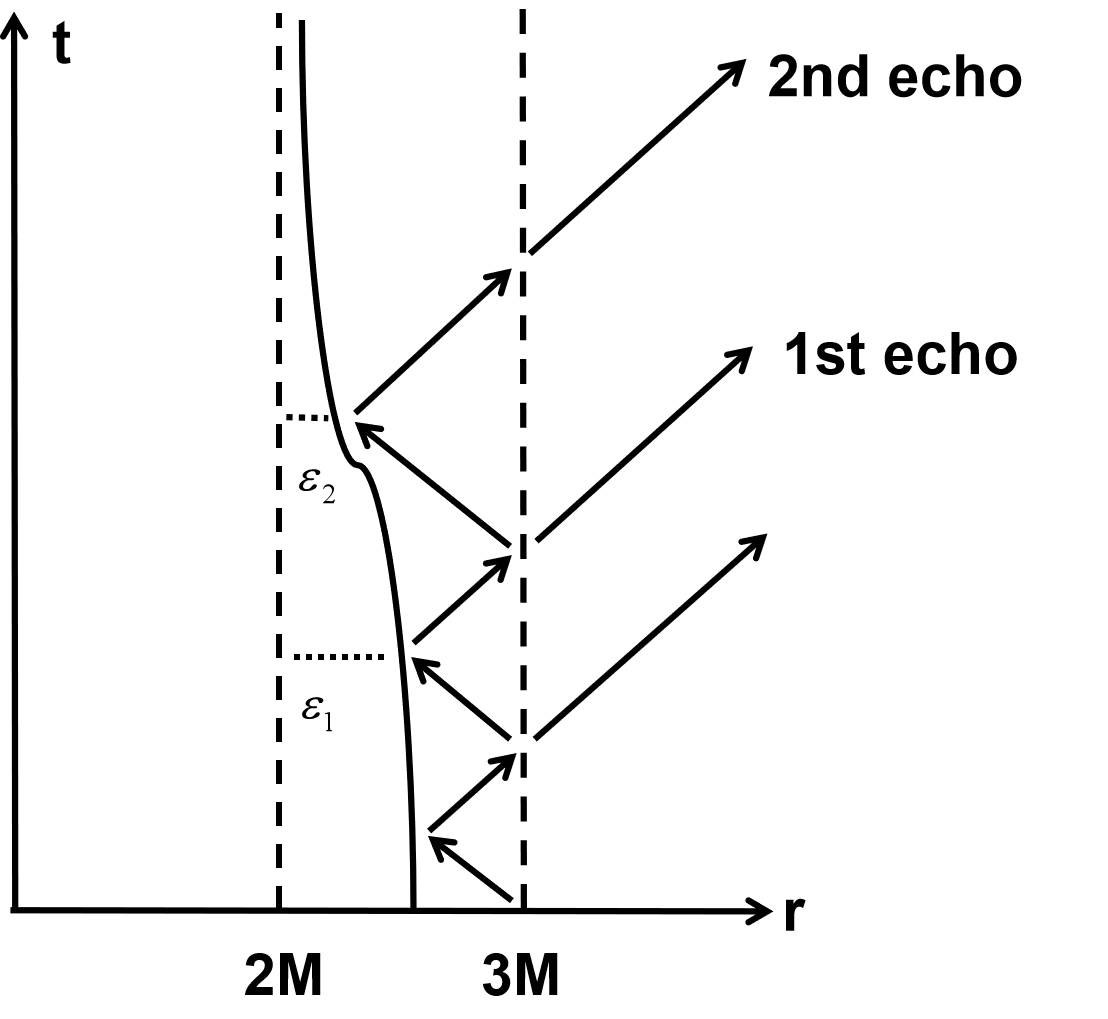}
\caption{Spacetime diagram of a gravitational wave scattering on a physical black hole on $\{t,r\}$ plane. The left curved represent the star surface while the right vertical line represent the photon sphere.}
\label{fig:schem}
\end{center}
\end{figure}

In the following we will consider the scattering of a Gaussian shaped gravitational wave burst. For estimation we neglect the width of the centrifuge potential and the burst itself at first. So the propagation of the wave burst can be studied with geometry optics. Note that the gravitational wave travels at the speed of light so it will surely catch up with the collapsar surface and get bounced back, as sketched in Fig.(\ref{fig:schem}).

\subsubsection{time-delay}\label{sec:timedelay}

Now we consider an ingoing pulse hitting on the photon sphere at $t=0$. A portion of the pulse will be reflected by the centrifuge potential and travel outward along a light-like geodesic. To all probes or observers located far away $R\gg 3M$, the estimated time at which the first signal will be recorded at is
\begin{equation}
\begin{split}
t_{1}&=\int_{3M}^{R}dr\frac{1}{1-\frac{2M}{r}}\\
&=(R-3M)+2M\ln\frac{R-2M}{M}.
\end{split}
\end{equation}

The transmitted part of the pulse will keep going inward until it hit the collapsar surface and get bounced back. Part of this signal will penetrate the centrifuge potential and recorded by the observer as the first echo at time
\begin{equation}
\begin{split}
t_{2}&=t_{1}+2\int_{3M}^{2M+\varepsilon_{1}}dr\frac{1}{1-\frac{2M}{r}}\\
&=t_{1}+2(M+2M\ln\frac{M}{\varepsilon_{1}}),
\end{split}
\end{equation}
where $\varepsilon_{1}$ is the distance between the Schwartchild radius and the collapsar surface when it is hit on by this pulse, as shown in Figure \ref{fig:schem}.

For two neighboring echos, the difference between their recorded time is
\begin{equation}\label{eq:deltat}
2(\int_{3M}^{2M+\varepsilon_{n-1}}dr\frac{1}{1-\frac{2M}{r}}-\int_{3M}^{2M+\varepsilon_{n}}dr\frac{1}{1-\frac{2M}{r}})=2(\varepsilon_{n}-\varepsilon_{n-1}+2M\ln\frac{\varepsilon_{n}}{\varepsilon_{n-1}})
\end{equation}
Since the physical black hole has already entered its very late time of the collapse, $\varepsilon_{i}\to 0,i=1,2,\cdots$. In this case  Eq.(\ref{eq:deltat}) can be approximated by $4M\ln(\frac{\varepsilon_{n}}{\varepsilon_{n-1}})\sim-4M\frac{\varepsilon_{n-1}-\varepsilon_{n}}{\varepsilon_{n-1}}$.

\subsubsection{Redshift}

We now assume that the wave burst is monochrome. In the Schwartzchild coordinate, its wave vector can be written as
\begin{equation}
K^{\mu}=\{\frac{\omega}{1-\frac{2M}{r}},\omega,0,0\}
\end{equation}
where $\omega$ is the observed frequency by static observers located on $r$. If the world line of the incoming wave and collapsar surface intersect at $(t,r)$, then to the co-moving observer fixed on the collapsar surface, the frequency would be
\begin{equation}
\omega'=-g_{ab}K^{a}(\frac{\partial}{\partial\tau})^{b}=(\dot{t}+\frac{\dot{r}}{1-\frac{2M}{r}})\omega.
\end{equation}
The over dot here denotes differentiations with respect to the collapsing surface's proper time. After reflection the outgoing wave vector in the Schwartzchild coordinate can be written in the form
\begin{equation}
K'^{u}=\{\frac{\omega''}{1-\frac{2M}{r}},-\omega'',0,0\}
\end{equation}
By the relation $\omega'=-g_{ab}K'^{a}(\frac{\partial}{\partial\tau})^{b}$, we know
\begin{equation}\label{eq:doppler}
\omega''=\frac{\omega'}{\dot{t}+\frac{\dot{r}}{1-\frac{2M}{r}}}=\frac{\dot{t}+\frac{\dot{r}}{1-\frac{2M}{r}}}{\dot{t}-\frac{\dot{r}}{1-\frac{2M}{r}}}\omega.
\end{equation}
From Eq.(\ref{eq:doppler}) we see that there is a huge redshift in the reflected signal if the speed of the collapsar surface reaches $\mathcal{O}(1)$ fraction of the local speed of light.

\subsection{The waveform and echos}

\subsubsection{Setups}

To obtain the gravitational waveform, we will numerically simulate the propagation of wave bursts according Eq.\eqref{eq:eom} with the aforementioned boundary condition. In our simulations, the mass is normalized to $M=1$, and the spatial infinity is placed at $x=100$. In mathematics, we known that the moving Dirichelet boundary condition \eqref{eq:mvb} is indistinguishable from a delta function barrier with infinite height placed in the same place. In our program, we use a thin Gaussian function shaped potential to do this job. So the total potential in the master equation Eq.\eqref{eq:eom} becomes
\begin{equation}
V(x)=V_{l=2}(r)+V_{B}(x,X(t)).
\end{equation}
The boundary potential $V_{B}(x,X(t))=h e^{-\frac{(x-X(t))^{2}}{d}}$ here is the thin Gaussian barrier used to model the collapsar surface and the parameters are chosen as $h=100$ and $d=0.1$, respectively.

The initial waveform is chosen as
\begin{equation}
\psi(0,x)=e^{-\frac{(x-x_{0})^{2}}{\sigma^{2}}},\partial_{t}\psi(0,x)=0,
\end{equation}
which is centered at $x_{0}=10$ and broadened to width $\sigma=1$. What we are going to plot is the waveform recorded by observers placed at $x=70$. The motion of the collapsar surface is chosen as
\begin{equation}\label{eq:xt}
X_{\varepsilon}(t)=-30-\frac{t}{4},
\end{equation}
corresponding to the collapsar surface contracting at $1/4$ times of the speed of light. The speed of the collapsar surface depends on the collapsing material's equation of state, so is not uniquely determined. Our choice here is a result of taking balance between the need of clear demonstration and saving of computational resources.

\subsubsection{Waveform}

\begin{figure}
\begin{center}
\includegraphics[width=.79\textwidth]{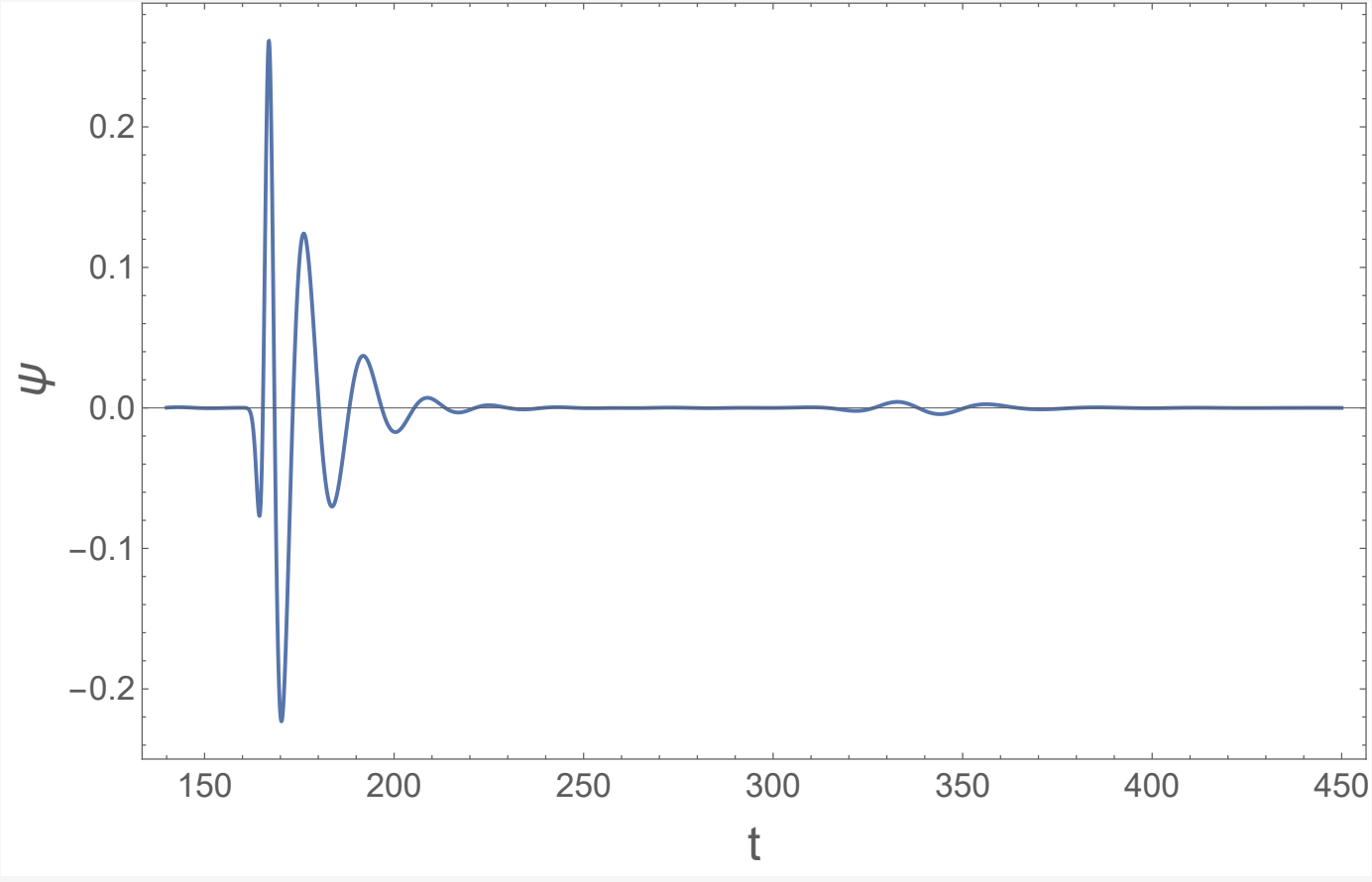}
\caption{The echos from a Gaussian wavepacket scattering off a physical black hole. Two echos can be seen in this waveform, one around $160-240$ and another one around $300-390$.}
\label{fig:mbwaveform}
\end{center}
\end{figure}

The projected waveform is presented in Figure \ref{fig:mbwaveform}, where the duration before the echos is discarded. The portion of the initial wave burst that penetrated the centrifuge potential will catch up with the collapsar surface and be reflected. After that it will be scattered by the centrifuge potential again. Part of this signal will transmit through the potential and gives rise to the first echo in the waveform recorded by the gravitational wave detector at $t\sim160-240$. The other part of the signal scattered back by the centrifuge potential will repeat the same process and form a series of echos. In Figure \ref{fig:mbwaveform}, we can see the second echo around $t\sim300-390$.

\begin{figure}
\begin{center}
\includegraphics[width=.79\textwidth]{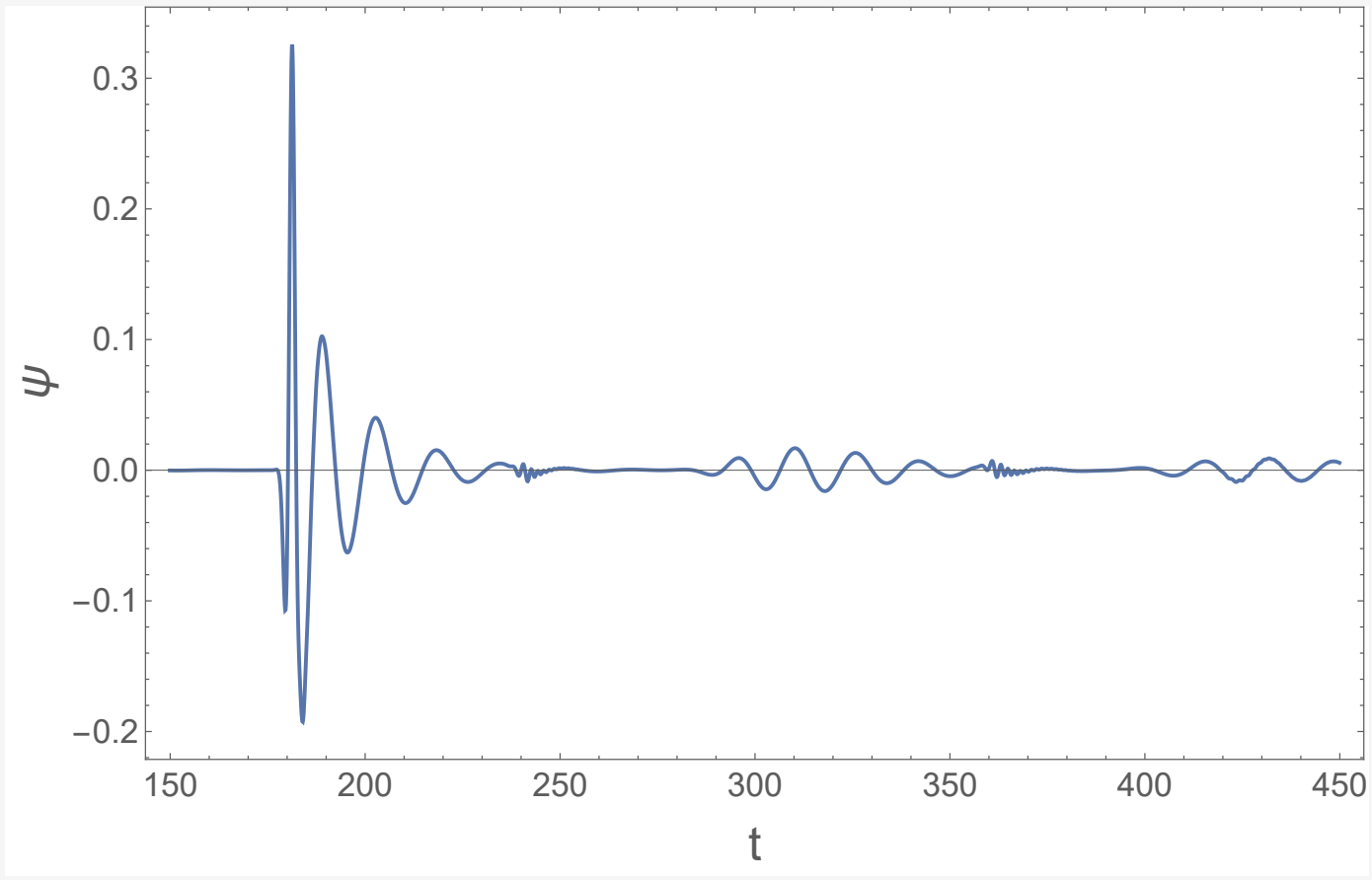}
\caption{The echos from a Gaussian wavepacket scattering off an ECO. Three echos can be seen in this waveform, the first lies around $170-270$, the second lies around $280-390$ and the last one starting from $390$. The three echos are equally spaced with interval of $110$ time units.}
\label{fig:fbwaveform}
\end{center}
\end{figure}

We put the waveform of echos from an ECO \cite{jcap,D94,na} in Figure \ref{fig:fbwaveform} for comparison, where all other parameters are chosen identically as above except for the star surface location. For the ECO in Figure \ref{fig:fbwaveform}, the star surface is statically located at $x=-50$. The figure shows three distinct echo signals, the first two are roughly located at $t\sim170-270$ and $t\sim280-380$, while the third starts around $t\sim390$ and persists till the end of the simulation.

Comparing the two waveforms we immediately see two differences. The first is that echos from physical black holes are not equally spaced, while those from ECO are equally spaced. Among the three echos in Figure \ref{fig:fbwaveform} we see that the two intervals between neighboring 3 echos are both roughly 110 time units. In Figure \ref{fig:mbwaveform}, the interval between the first two echos is about 140. The third echo should appear at around $t\sim440$ if the echos are equally spaced. However, it does not manifest till the end of the simulation at $t=450$. This confirms our estimation in Sec.\ref{sec:timedelay}. The second difference is the strength of the echos. At this stage, we can not tell if there is difference between the strength of first echos from the two figures, but we can easily see that the second echo in Figure \ref{fig:mbwaveform} is weaker than its counterpart in Figure \ref{fig:fbwaveform}. The reason behind this phenomenon is as follows. The ingoing wave receives Doppler red shift when reflected by the collapsar surface. When it tries to penetrate the centrifuge potential, the low frequency components will be stopped further. Thus the echos from the physical black holes looks fainter than those from ECO. In the next part we will dive deeper on this issue.

\subsubsection{Spectra}

\begin{figure}
\begin{center}
\includegraphics[width=.89\textwidth]{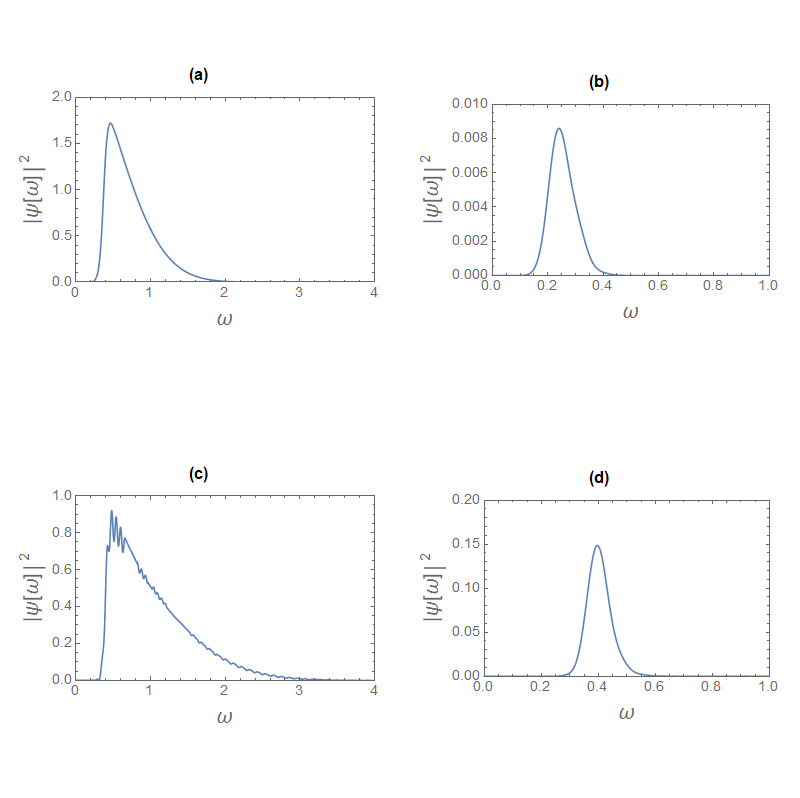}
\caption{Subdiagrams (a) and (b) display the absolute value of the Fourier spectrum of the first echo and second echo from physical black holes. Those from ECO are displayed in subdiagrams (c) and (d) respectively.}
\label{fig:spec}
\end{center}
\end{figure}

This part applies Fourier analysis to echos in Figures \ref{fig:mbwaveform} and \ref{fig:fbwaveform} to examine their frequency spectrum. We see that each of the echos are isolated inside a specific time interval. Hence the Fourier transformation of the echos can be computed via \cite{jcap}
\begin{equation}
\psi(\omega)=\int_{t_{i}}^{t_{f}}dt\psi(t)e^{i\omega t},
\end{equation}
where $t_{i}$ and $t_{f}$ are the beginning and the ending of the time interval of a specific echo signal, respectively. The spectrum of echos from physical black holes are presented in Fig.\ref{fig:spec}(a) and (b), while that from ECO are presented in Fig.\ref{fig:spec}(c) and (d). We see that the peak frequency of the first and second echos from ECO are both located at roughly $0.4$. For the physical black hole, the peak frequency of the first echo is about $0.4$, while that of the second echo is about $0.25$. This is a quantitative demonstration of Doppler shift from the reflection on the collapsar surface. Comparing Figure \ref{fig:spec}(b) and \ref{fig:spec}(d), we easily see that the second echo from physical black holes is fainter than that from ECO. Examine Figure \ref{fig:spec}(a) and \ref{fig:spec}(c), since the area enclosed by the power spectrum and horizontal axis represents the energy of the echo signal, we can see that the first echo from physical black hole is also weaker than that from ECO, as expected. The reason this difference is unclear in the time domain waveform Figure \ref{fig:mbwaveform} and \ref{fig:fbwaveform} is that, the radial speed of the collapsar surface is set to be relatively small in order to see desired echo signal in our computational resources, thus yielding a smaller Doppler red shift factor.

Figure \ref{fig:spec}(c) and (d) give a demonstration on the tendency of the centrifuge potential barrier reflecting the low frequency components, as we mentioned in Sec. 3.2.2. Comparing Figure \ref{fig:spec}(c) and (d) we see that the high frequency component in the first echo is absent in the second echo. Similar things also happen in Figure \ref{fig:spec}(a) and (b) but not so obviously. This can be understood the following way. We assign a reflection factor $R(\omega)$ and a transmitting factor $T(\omega)=1-R(\omega)$ to the centrifuge potential. For a specific potential with finite height, $T(\omega)$ has an overall tendation to decrease with lower frequency. This explains what we see in Figure \ref{fig:spec}(c) and (d). That is, the high frequency component of the first echo is transmitted outside through the centrifuge potential and the low frequency component is reflected back and left for the second echo.

\section{Conclusion and discussion}

Astrophysical black holes forms through gravitational collapse. The time dilation mechanism of general relativity  implies that in the time definition the black hole is defined and measured by outside observers, the system is described by a dynamical and horizonless geometry, which asymptotically developes but never truly implements a successful horizon. The reason behind the privilege of Schwartzchild time in black hole phenomenology lies in that all observational signal comes from probes living outside the black holes, whose entire history lies only in the domain of such a time definition. With the aim of establishing a testable signal to distinguish the physical black holes from mathematical ones, we considered the gravitational wave echos from a gravitational burst scattering off a physical black hole. We found that such echos are inevitable consequences for physical black holes and have distinct features comparing with echos investigated in other scenarios. The differences can be seen in both the time and frequence domain. Specifically, the echos from physical black holes are not equally spaced, and there is a huge Doppler redshift among the neighboring echos. Our results indicate that the presence of gravitational wave echos does not necessarily implies the existence of new physics. Rather, it is comprehensible within the framework of classical general relativity.

The original signal stimulating echos arises from the ring-down motion of a binary merger system. A natural question arises here, are there echos produced druing the inspiral and merger motion of the binary system? The answer is yes. During all phases, the gravitational wave propagates all directions. Most of it will radiate freely to infinity while a small fraction will hit the binary system surfaces and bounce back, thus forming echos. It is noteworthy that the feature of echos from these phases could be drastically different from our results even for binary systems composed of two non-spinning black holes. The key reason is, during the inspiral and merger phases, the matter surface providing the reflective condition is far from spherically symmetric.

\section*{Acknowledgements}

This research was funded by the National Natural Science Foundation of China under Grant No.11785082.

\end{document}